# Molecular coupling of light with plasmonic waveguides

Anton Kuzyk,[1] Mika Pettersson*,[2] J. Jussi Toppari,[1] Tommi K. Hakala,[1] Hanna Tikkanen,[2] Henrik Kunttu,[2] Päivi Törmä[1]

[1]Nanoscience Center, Department of Physics, P.O. Box 35, FIN-40014, University of Jyväskylä, Finland
[2]Nanoscience Center, Department of Chemistry, P.O. Box 35, FIN-40014, University of Jyväskylä, Finland
*Corresponding author: mijopett@cc.jyu.fi

**Abstract:** We use molecules to couple light into and out of microscale plasmonic waveguides. Energy transfer, mediated by surface plasmons, from donor molecules to acceptor molecules over ten micrometer distances is demonstrated. Also surface plasmon coupled emission from the donor molecules is observed at similar distances away from the excitation spot. The lithographic fabrication method we use for positioning the dye molecules allows scaling to nanometer dimensions. The use of molecules as couplers between far-field and near-field light offers the advantages that no special excitation geometry is needed, any light source can be used to excite plasmons and the excitation can be localized below the diffraction limit. Moreover, the use of molecules has the potential for integration with molecular electronics and for the use of molecular self-assembly in fabrication. Our results constitute a proof-of-principle demonstration of a plasmonic waveguide where signal in- and outcoupling is done by molecules.

## 1. Introduction

Surface plasmon polaritons (briefly, surface plasmons (SP)) offer fascinating prospects for understanding and exploiting phenomena related to nanoscale confinement of light. SPs are coupled modes of electromagnetic field and free electrons in metal. They can be considered as two-dimensional light which is bound to a metal-dielectric interface and can be confined to dimensions much smaller than wavelength of the light in free space. Propagation and scattering as well as energy transfer with molecules are all modified by the subwavelength confinement of these optical fields [1,2]. Ultimately, SPs may provide the path to integrated electrical and optical circuits for information processing at the nanoscale. The use of molecular components in downscaling electrical circuits has been intensively studied in the context of molecular electronics [3]. On the other hand, the enhanced interaction of molecules with SPs is widely applied, e.g. using surface enhanced Raman scattering (SERS) with tremendous sensitivity [4]. However, combining molecules with SP components that are suited for circuitry, such as waveguides, is mostly an unexplored territory. In this article, we investigate the concept of a plasmon waveguide which is connected to the environment via molecules.

Development of lithographic techniques has enabled significant progress in guiding and manipulating SPs in nanoscale metallic structures: guiding of SPs in metal stripes [5], dielectric stripes [6], nanowires [7], and V-grooves over up to tens of micrometers has been demonstrated [8], as well as optical elements such as interferometers, resonators, beam splitters and Bragg reflectors [7-9]. SPs can be used to enhance fluorescence of molecules or to enhance Raman scattering via the SERS effect [10]. These phenomena are important for many analytical applications, especially in biochemistry and biomedical research. On the other hand, fluorescent molecules can be used to couple light from far field to plasmons without the need for any special arrangements as is the case when light is coupled directly to plasmons, for example, by using a prism in a Kretschmann geometry [2]. Coupling light to SPs via molecules offers another significant advantage: if molecules can be arranged in a controlled way on a metal surface in a small, well defined, region the initial launching of SPs can be done in an area much smaller than the diffraction limit, determined only by the size of the molecular ensemble. In this respect, lithographic methods or molecular self-assembly can offer methods for achieving nanoscopic dimensions. Molecules can also be used to couple SPs to far field light as is done in visualization of SPs by using fluorescent molecules [9]. Therefore, by a design of molecular structures on metal surfaces it is possible, in principle, to realize a full cycle from far field light to SPs and back to far field light, thus providing a versatile method to couple light into SPs and obtain information on the plasmonic processes. Such a cycle may be envisioned to be used for realizing coupled wires in "molecular plasmonics" information processing. As another fascinating application, SPs can be used for molecular energy transfer over large distances. Such energy transfer can find applications, for example, in artificial light-harvesting structures or organic light emitting diodes [11]. Andrew and Barnes demonstrated energy transfer over more than 120 nm through a thin silver film between donor and acceptor molecules [11]. This distance is about an order of magnitude larger than achieved by Förster type of resonance energy transfer (FRET) but still highly efficient unlike normal radiative transfer through free space propagation. Plasmonic energy transfer has also been observed on two-dimensional surfaces for silver nanoparticle aggregates [12,13]. In this article we demonstrate such energy transfer on a two-dimensional lithographically patterned metal surface, over distances of ten micrometers. Concerning the information processing and the molecular energy transfer applications, the feasibility of fabricating a combined molecular and plasmonic device is a key requirement. For the positioning of the molecules, we introduce a scalable lithographical method which allows immobilization of several different types of molecules at predetermined locations.

The structure of this article is the following: First the fabrication of the structures and the measurement methods are described in Section 2. Then, in Section 3, we confirm that the propagation of light in the waveguide is of plasmonic nature by analyzing images in Fig. 2 and Fig. 4. Reference structures are used for ruling out the contribution of free space propagation, and dynamics of the process (Fig. 5) shows that the input coupling is of molecular origin. Finally, spectral information, Fig. 4, is used for distinguishing the contributions of the donor (input coupling) and acceptor (output coupling) molecules, which demonstrates the full cycle of coupling light from far field to propagating plasmons and back via molecules. The demonstration of the full cycle, together with the scalable fabrication method, constitute the main results of this article.

## 2. Sample fabrication and experimental methods

For positioning the donor and acceptor molecules we propose the use of a lithographic method where the molecules are embedded into SU-8 polymer resist (Microchem SU-8 2000 series) which is consequently processed by e-beam lithography, leaving the exposed area on the chip: the method is scalable and can be repeated in cycles to position several different molecules. Note that use of a positive resist like PMMA, where areas unexposed to the e-beam are left on the surface, is not feasible because a polymer area positioned in the first cycle would be harmed by the exposure during a subsequent cycle. Relevant to our approach is that, quite recently, dye-doped SU-8 has been used for fabricating microscale polymer laser [14], and the know-how developed thereby can now be applied in the plasmonics context. Our results show that the dye molecules maintain their fluorescent properties in e-beam processing, which is important since e-beam lithography allows efficient downscaling. Before positioning the molecules, the 5 μm wide plasmonic silver waveguides were fabricated by regular e-beam lithography and lift-off techniques on a top of an ITO coated glass, as in our earlier work [15]. The waveguides consist of a thin adhesion layer (2 nm) of titanium and on top of that a 100-130 nm thick evaporated layer of silver (evaporation rate 3-4 nm/s in a UHV chamber). The dye-SU-8 solution was spin coated (3000 rpm) on top of the waveguide. In the used dye-SU-8 solutions the mass percentage of the donor molecule Coumarin 30 (C30), of the dry SU-8 resin weight, varied from 2.5 % to 11 %, and the mass percentage of the acceptor Rhodamine 6G (R6G) varied between 2.5 % and 7.9 %. The thickness of thin layers was controlled by adjusting SU-8 polymer resin concentration by dilution with a suitable amount (as specified for the Microchem SU-8 2000 series) of cyclopentanone. The actual thickness was verified by AFM. The immobilization of the dye molecules by e-beam lithography of the SU-8 resist involved a precise alignment step using marks created at the same process step as the waveguide. Finally, the patterning was done on the areas where the dye molecules were intended and the rest of the surface was cleaned using SU-8 developer (Microchem).

The layer structure of a sample is shown schematically in Fig. 1(A). Red squares correspond to the acceptor molecules embedded in SU-8 and cyan area to SU-8 with the donor molecules. The long silver stripe is the waveguide for plasmon propagation and the squares on the sides provide a reference measurement to rule out the contribution from free space propagation. An AFM image of a typical sample is shown in Fig. 1(B).

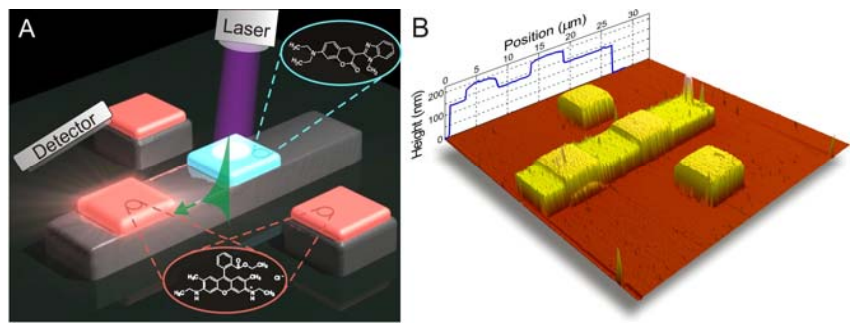

Fig. 1. A) Schematic presentation of the structure of the samples and the measurement (see text for details). B) Measured AFM image of a typical structure. The height profile along the center of the waveguide is shown on the left. The size of the square regions is 5 x 5 μm2.

Simultaneous excitation of the donor area and collection of light from the acceptor areas was achieved by carrying out the measurements in a confocal microscope equipped with a dual scanner (Olympus Fluoview-1000). The first scanner controlling the diode laser at 405 nm was used to locally excite the donor molecules (C30) in the prepared microstructures. The scanning mode was "tornado", i.e. spiral-like motion inside the selected region (circle with a diameter of 3-4 μm). The laser power was kept low (10-100 μW) in order to reduce bleaching of the dye molecules to allow long enough measurement times (of several minutes). The second scanner was used to collect the emission from the whole sample point by point, resulting in a confocal image. It should be emphasized that the second scanner was not used for excitation but only for collection. The excitation laser was filtered by a dichroic mirror when the detection bandpass was near the excitation wavelength. Otherwise, a beam splitter was used allowing collection of spectra which were not modified by filters. The emission spectra were recorded with the same instrument by scanning the spectral region in 3 nm steps of 6 nm bandpass. A 100 X air objective (NA = 0.95) was used for excitation and collection and the scanning speed for excitation and measurement was 10 ms/pixel and 40 ms/pixel, respectively. The size of the measurement area was 800 X 800 pixels$^2$ corresponding to the dimensions of 42 X 42 μm$^2$.

## 3. Experimental results

### 3.1 Plasmon propagation

The plasmon propagation is demonstrated by confocal images where scattering from several places along the metal stripe is observed whereas the reference areas, with no metallic connection to the excitation area, show very little or no signal. To visualize the structure, Fig. 2(A) shows an image of a sample, where a donor is positioned in the center of the structure in a 5 x 5 μm$^2$ area and the acceptor is placed in the upper, right and left arms of the cross in regions with the dimensions of 5 x 5 μm$^2$, while the lower arm has no dye on it. The thickness of the donor layer was 50 nm in all the samples, and in this structure, the thickness of the acceptor layer is 60 nm.

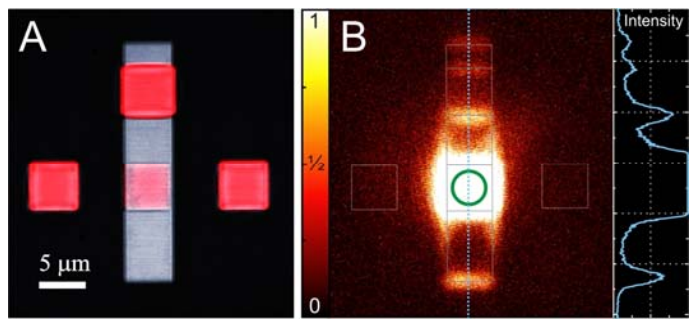

Fig. 2. Confocal microscope images of the plasmonic waveguides. A) An image taken with a single scanner setup, by scanning the sample with the excitation light and simultaneously collecting data in two channels: First channel, the red color corresponds to a collection band that includes emission from the acceptor (R6G; the left, right and up squares) and partly from the donor (C30, the middle square). Second channel, the grey-blue corresponds to the reflection image and shows the metal waveguide. B) An image taken with a dual scanner setup where the donor molecules are excited with 405 nm (The green circle corresponds to the area of excitation) and emission with a bandpass of 520 – 620 nm is collected with a second independent scanner. The contour of the waveguide is overlaid with the intensity map for easy comparison. The thickness of the acceptor layer is 60 nm and the thickness of the donor layer is 50 nm. The intensity profile along the blue dotted line, indicated in the figure, is shown on the right.

The upper and the lower arms are connected to the center by a silver stripe providing a plasmonic waveguide channel while the disconnected squares on the sides act as references. The image of the structure in Fig. 2(A) is taken with 488 nm excitation. The red color shows the signal collected with a 520 – 620 nm bandpass transmitting most of the emission intensity of the acceptor and also part of the emission spectra of the donor. The grey-blue color shows a superimposed reflection image where the metal stripe is visible. The Fig. 2(B) shows an image taken with the dual scanner configuration by exciting the region of the donor with 405 nm laser and simultaneously recording the image with the second scanner. Again, the collection bandpass is 520 – 620 nm. The signal in the region of the donor is saturated due to direct excitation of this region. A weaker signal is observed in different regions of the plasmonic channel while there is essentially no signal from the reference regions, i.e., left and right arms. This clearly indicates that the observed signal originates from SPs. The absence of signal in the arms without the plasmonic channel rules out free space propagation as the source of excitation energy transfer.

In order to further check that the propagation is due to SPs we prepared samples with varying length (7.5–22.5 μm between the donor edge and the stripe end) and measured the intensity of scattered signal at the end of the stripe. The dependence of signal intensity on the length showed exponential behavior with a characteristic propagation length varying (due to silver quality) from 7 to 10 μm between the sets of samples. Each set of samples was fabricated during the same process to the same chip and included all the measured stripe lengths. Results from one set are shown in Fig. 3. The decay was also determined by measuring the scattered intensity along the stripes yielding similar propagation lengths. These values are in agreement with the obtained propagation length of ~20 μm for extended silver at 514 nm (silver-air interface) [5]. For 633 nm wavelength and for a stripe width of 5 μm the propagation length is approximately half from the value in extended silver. Using this relation gives a perfect agreement between our data and that of Lamprecht et al [5].

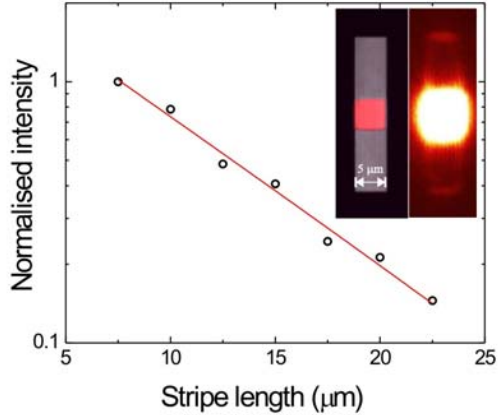

Fig. 3. Intensity of the scattered signal at the end of the stripe as a function of the stripe length (black circles). The red line is the fitted exponential decay yielding propagation length of 7.7 μm. Inset: Confocal microscope images of one of the measured waveguides taken with the single scanner and the dual scanner setups (see Fig. 2 and text). This waveguide has 10 μm from the edge of the donor (in the middle) to the end of the stripe, where signal from scattered SPs is clearly visible in the two scanner image.

The edges of the polymer areas and of the metal stripe provide efficient scattering centers for the propagating SPs. Closer inspection of the signal in Fig. 2(B) in the region of the acceptor molecules shows that most of the intensity arises from the front edge (looking from the donor) of the acceptor region, and to a lesser extent from the back edge of the acceptor region or from the far edge of the plasmonic channel. In the lower arm where there is neither dye nor polymer, the emission is observed from the far edge of the plasmonic channel. These observations point out that the SPs emit to the far field mainly at the regions which provide a suitable geometry for scattering such as sharp edges and refractive index steps.

### 3.2 Incoupling of plasmons by molecules

The evidence that the excitation of plasmons is due to molecule emission is given by the spectral information from the emissions as presented in Fig. 4. This figure gives evidence that surface plasmon coupled emission (SPCE) of donor was observed in our experiment and also that we indeed realize a full cycle from far field to SP propagation and back to far field via molecules, as discussed later. Fig. 4 shows a single plasmonic channel with the donor region in the center and the acceptor region in the upper arm. The thickness of the donor layer is 50 nm and the thickness of the acceptor layer is 550 nm. The lower arm contains no acceptor molecules but is otherwise identical to the upper arm. The spectra were taken using the dual scanner setup as described in the beginning of this article. Let us first discuss the lower arm. The strongest emission signal can be observed at the back edge [16] of the region with SU-8 resist (this region is the same as the acceptor region in the upper arm but without dye molecules). The spectrum of the signal is shown in the inset (c) of Fig. 4(B). The spectrum is nearly identical to the spectrum of C30 measured by a direct excitation (shown with a blue line). We interpret the emission shown in inset (c) as surface plasmon coupled emission (SPCE) originating from coupling of the emission of the donor to the SPs and radiating back to far field on a scattering center far away. Here the SP scatters light to the far field at the edge of the SU-8 layer which serves as a scattering center. This SPCE of the donors happens also in the upper arm of the structure, although there the acceptor contribution is visible as well, as discussed in detail below. In fact, if no efficient scattering centers are available, most of the emission is lost by dissipation into heat [10]. Lakowitz et al. have recently reported a series of investigations on SPCE [17-20].

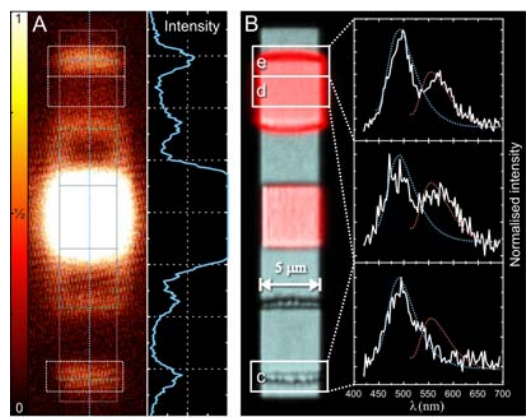

Fig. 4. Confocal microscope images of a plasmonic channel. A) An image obtained by a dual scanner measurement where the excitation is only on the donor area in the middle and the collection is scanned over the sample, showing plasmon propagation as the scattered light along the metal stripe. Description of how the images are taken can be found in the caption of Fig. 2(B) and in the text. The center rectangular region (see Fig. 4(B)) contains the donor molecules while the rectangular region in the upper arm contains the acceptors. The rectangular region in the lower arm is an SU-8 structure otherwise analogous to the upper arm except that it does not contain any dye. The intensity profile along the blue dotted line is shown on the right. B) The spectra measured at different regions under excitation of the donor with 405 nm laser are shown in the insets (c) – (e). The reference spectra of the donor (blue) and acceptor (red) measured by direct excitation are shown for comparison. That the donor spectra is observed far away from the excitation region is a signature of surface plasmon coupled emission (SPCE), and the acceptor spectra demonstrate molecular energy transfer via surface plasmons over the ten micrometer long distance.

They excited dye molecules on top of a thin metal film which was coupled to a prism and observed directional emission through the prism. Their arrangement corresponds to a reverse Kretschmann geometry where excited SPs radiate to the far field at a certain angle. Our observations of the donor SPCE are related to those of Lakowitz et al. except that in our case the excitation and emission are spatially separated by about 10 μm, the excitation area is in the microscale, and the emission is via scattering from a microscale refractive index step rather than via a prism. That the initial excitation is coupled to plasmons via donor molecules and not because of scattering from the surface imperfections can be deduced also from an analysis of the observed bleaching. The donor molecules bleach in a timescale of a minute under excitation, despite the reduced intensity (~20 μW). The observed signal from the acceptor region decays with the same rate, which rules out scattering from the surface as a possible origin of the SP excitation. Fig. 5 shows decay curves of the emission collected from the regions defined in Fig. 4(B). The curves were obtained by exciting continuously the donor region with the 405 nm laser and recording the emission from specified regions with a second scanner. The decay of intensity was observed to follow biphasic kinetics in regions (c) – (e) with characteristic time constants of $t_1 = 11 - 16$ s and $t_2 = 62 - 73$ s. The directly excited C30 yielded a similar curve with a time constant $t_1$ of 14.6 s and additionally a short component with $t = 2.3$ s. This short component was not observed in the other measurements because for those regions the measurement time step was longer (5 s) than the time constant. The similarity of the decay constants in all the regions indicates that all the emissions in the regions (c) – (e) originate from the same process, namely coupling of the excited donors to SPs. Moreover, the decay of the signals is caused by bleaching of the donor molecules, verifying the molecular origin of the SP coupling.

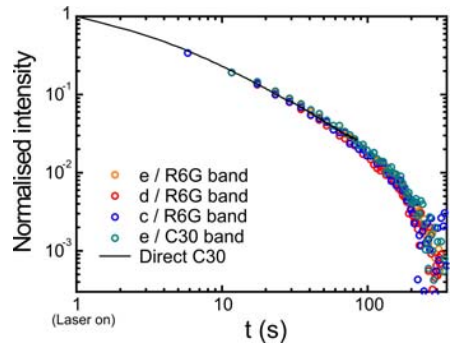

Fig. 5. Normalized decay curves of the emission intensity measured at different regions (c) – (e) defined in Fig. 4, with detection bandpass corresponding to the maximum intensity of the donor (C30) or the acceptor (R6G). Each curve is measured from a different identically prepared sample. The curve labeled "direct C30" corresponds to the decay curve of directly excited donor, and other signals correspond to emission induced by excitation via SPs.

The fact that the same time constant was observed from the SPCE of the donor without acceptor and also from acceptor emission proves that bleaching of acceptor does not play a major role. We also made reference experiments where the excitation laser was directed to the metal surface. In this case, no significant signal along the metal stripe was observed, showing that the metal surface is sufficiently smooth so that incoupling of plasmons via surface roughness is not important.

*3.3 Outcoupling of plasmons by molecules*

Outcoupling of the plasmons via molecules is shown by the spectra in Fig. 4(B), insets (d) and (e). The polymer matrix in the upper arm of the structure contains acceptor molecules. The strongest emission signal appears at the back edge of the rectangular region but some intensity is also observed in the center of the acceptor region. In both regions, the spectrum is composed of spectra of both the donor and the acceptor as can be seen by comparison with the emission spectra of directly excited donor and acceptor. The relative contribution of the acceptor is strongest in the spectra measured at the center of the acceptor region, but at the edge its contribution is also substantial. The *donor emission* in the upper arm can be identified as SPCE, as discussed above. The absolute intensity of the total emission in the center of the acceptor region (d) is lower than at the edge (e) as can be seen from Fig. 4(A). In the center of the acceptor area (d) the donor SPCE could originate from scattering of SPs from the imperfections of the polymer area. The remaining interesting questions are what are the mechanisms of the acceptor excitation, and acceptor emission to the far field? In principle, the acceptors could be excited by the donor SPCE light but this effect should be small because the donor SPCE light originates mostly from the edges of the acceptor area and is scattered to the full half space. It is thus plausible that most of the acceptor excitation is caused directly by SPs. Then what about the *acceptor emission*? It could be either direct emission to the far field from the molecules excited by SPs, or SPCE in the sense that the acceptors, after being excited by SPs, may emit back to plasmons which consequently scatter from the edges of the polymer region. We cannot unambiguously determine from the present data whether the process is direct emission or SPCE, but based on the fact that the acceptor emission is stronger at the edge of the acceptor region (also evident from Fig. 4(A) and from the relative strength of the donor and acceptor emission) we find the most plausible explanation that the majority of the emission from the acceptor is SPCE. It is reasonable to expect that those molecules which are excited by the SPs can also be coupled back to SPs (now at the acceptor emission frequency). Therefore, one can expect to observe considerable SPCE at the scattering centers. In any case, the observation of the acceptor emission clearly shows that the initial excitation by 405 nm light is coupled to the SPs via donor molecules, followed by propagation of SPs

over ~10 μm and excitation of acceptors, thus demonstrating molecular coupling and plasmonic energy transfer

Finally, we consider the efficiencies of different steps in the cycle from far field light to SP excitation and back to far field light. The present experimental method (confocal microscopy) cannot give quantitative data on efficiencies of different steps which is a topic of further investigations. However, some estimates can be done. We start by calculating the absorption in the donor layer. According to literature data Coumarine 30 has a molar absorption coefficient of 42740 $M^{-1}$ $cm^{-1}$ at 405 nm [21]. Using this value for a typical sample with a thickness of 50 nm and a concentration of 5 % (by weight) yields that 15.6 % of the excitation intensity is absorbed. In this estimation the optical path length is multiplied by a factor of two due to a reflecting metal surface. The quantum yield for fluorescence adds another factor of 0.67 for the efficiency [22]. In a favourable case the coupling efficiency to plasmons can be of the order of 70 % [10]. The propagation length of SP is taken from our measurement to be 10 μm yielding a factor of 0.368 for a 10 μm long stripe. At this point our estimation shows that about 2.7 % of the initial intensity has been transferred to the acceptor site. Next we calculate absorption in the acceptor layer. From the emission spectrum of Coumarine 30 and molar absorption coefficient data for Rhodamine 6G, [21] we can estimate that along the 5 μm long acceptor area, 72.3 % of the SP coupled emission is absorbed. Altogether, our simple estimation gives that about 2 % of the initial laser power is absorbed in the acceptor region. Of course there are many unknown factors which were not taken into account in this estimation, such as effect of SPs on the absorption and emission efficiencies, variation of fluorescence quantum yield with distance from the surface, scattering of light from the surfaces and scattering of SPs from the surface imperfections, emission of SPs to the substrate and extension of SP field out of the acceptor layer. Therefore, the obtained number represents rather a best case scenario but nevertheless a realistic one.

## 4. Conclusions

Our results show that molecules can be used to couple light into plasmonic waveguides in the microscale and the presented fabrication method allows scaling down to nanoscale without any fundamental restrictions. Downscaling is limited only by resolution and alignment accuracy of lithographic techniques. The method is applicable also to, e.g., chemically prepared silver nanowires [7]. However, for possible applications, the stability of the donors and acceptors should be improved, i.e. bleaching of the dye molecules under laser irradiation should be reduced. We have made preliminary tests with quantum dots which are generally more resistant to bleaching than organic dyes. Since bleaching of molecules is known to be enhanced near metal surfaces, we also made tests where a 40 nm spacer layer of SU-8 containing no molecules was placed between the metal and the SU-8 layer containing the donors. Approximately 50 % longer decay times were observed suggesting further studies of this effect. An interesting direction of investigation is also to use atomic ions doped in hard dielectric materials as was recently reported in the work by Verhagen et al [23,24]. where $Er^+$ ions implanted in fused silica were used for SP imaging. Although the efficiency of launching of light to the direction of the waveguide was sufficient for the present experiment, also this issue could be improved, c.f. for instance [25]. Additionally, incoupling of light into plasmons and outcoupling of plasmons into light can be maximized by structuring both donor and acceptor regions into a suitable grating.

In summary, we have demonstrated the use of fluorescent molecules as couplers between far field light and surface plasmons in lithographically fabricated microstructures. The fabrication method allows easy scaling of the dimensions to the nanoscale. We have observed energy transfer from the donor molecules to the acceptor molecules over distances on the order of 10 μm. We also observed surface plasmon coupled emission at the scattering centers at similar distances from the excitation spot. In previous literature, SP propagation on micro- and nanostructures has been studied widely, as well as interaction of molecules with SPs on planar surfaces. Our work combines these two approaches in a novel way into a proof-of-principle experiment on one of the fundamental concepts for molecular plasmonics: plasmon propagation in a waveguide with molecular input and output coupling.

## Acknowledgements

This work was supported by Academy of Finland (project numbers 117937, 118160, 115020, 213362) and conducted as part of a EURYI scheme award. See www.esf.org/euryi. A.K. thanks the National Graduate School in Nanoscience. We thank M. Kaivola and J. Lindberg for discussions and V. Marjomäki for help in confocal microscopy.

## References:

1. W. L. Barnes, A. Dereux, T. W. Ebbesen, "Surface plasmon subwavelength optics," Nature **424**, 824 – 830 (2003).
2. A. V. Zayats, I. I. Smolyaninov, A. A. Maradudin, "Nano-optics of surface plasmon polaritons," Phys. Rep. **408**, 131 – 314 (2005).
3. J. R. Heath and M. A. Ratner, "Molecular electronics," Phys. Today **56**, 43 – 49 (2003)
4. K. Kneipp, H. Kneipp, I. Itzkan, R. R. Dasari, M. S. Feld, "Ultrasensitive chemical analysis by Raman spectroscopy," Chem. Rev. **99,** 2957 – 2975 (1999).
5. B. Lamprecht, J. R. Krenn, G. Schider, H. Ditlbacher, M. Salerno, N. Felidj, A, Leitner, F. R. Aussenegg, J. C. Weeber, "Surface plasmon propagation in microscale metal stripes," Appl. Phys. Lett. **79**, 51 – 53 (2001).
6. B. Steinberger, A. Hohenau, H. Ditlbacher, A. L. Stepanov, A Drezet, F. R. Aussenegg, A. Leitner, J. R. Krenn, "Dielectric stripes on gold as surface plasmon waveguides," Appl. Phys. Lett. **88**, 094104 (2006).
7. H. Ditlbacher, A. Hohenau, D. Wagner, U. Kreibig, M. Rogers, F. Hofer, F. R. Aussenegg, J. R. Krenn, "Silver nanowires as surface plasmon resonators," Phys. Rev. Lett. **95**, 257403 (2005).
8. S. I. Bozhevolnyi, V. S. Volkov, E. Devaux, J.-Y. Laluet, T. W. Ebbesen, "Channel plasmon subwavelength waveguide components including interferometers and ring resonators," Nature **440**, 508 – 511 (2006).
9. H. Ditlbacher, J. R. Krenn, G. Schider, A. Leitner, F. R. Aussenegg, "Two-dimensional optics with surface plasmon polaritons," Appl. Phys. Lett. **81**, 1762 – 1764 (2002).
10. W. L. Barnes, "Fluorescence near interfaces: the role of photonic mode density," J. Mod. Opt. **45**, 661 – 699 (1998).
11. P. Andrew and W. L. Barnes, "Energy transfer across a metal film mediated by surface plasmon polaritons," Science **306**, 1002 – 1005 (2004).
12. J. M. Gunn, M. Ewald, M. Dantus, "Polarization and phase control of remote surface-plasmon-mediated two-photon-induced emission and waveguiding," Nano Lett. **6**, 2804 – 2809 (2006).
13. S. A. Maier, P. G. Kik, H. A. Atwater, S. Meltzer, E. Harel, B. E. Koel, A. A. G. Requicha, "Local detection of electromagnetic energy transport below the diffraction limit in metal nanoparticle plasmon waveguides," Nat. Mater. **2**, 229 – 232 (2003).
14. S. Balslev, A. Mironov, D. Nilsson, A. Kristensen, "Micro-fabricated single mode polymer dye laser," Opt. Express **14**, 2170 – 2177 (2006).
15. S. Tuukkanen, J. J. Toppari, A. Kuzyk, L. Hirviniemi, V. P. Hytönen, T. Ihalainen, P. Törmä, "Carbon nanotubes as electrodes for dielectrophoresis of DNA," Nano Lett. **6**, 1339 – 1343 (2006).
16. The fact that for different SU-8 thicknesses the strongest signals are observed at different points (front edge for 60 nm SU-8 region as in Fig. 2 and back edge for 550 nm SU-8 region as in Fig. 4) shows that the height of the polymer layer (refractive index step) is a crucial parameter in the reflection/scattering properties of SPs [2].
17. J. R. Lakowicz, "Radiative decay engineering 3. Surface plasmon-coupled directional emission," Anal. Biochem. **324**, 153 – 169 (2004).
18. I. Gryczynski, J. Malicka, Z. Gryczynski, J. R. Lakowicz, "Radiative decay engineering 4. Experimental studies of surface plasmon-coupled directional emission," Anal. Biochem. **324**, 170 – 182 (2004).
19. J. R. Lakowicz, "Radiative decay engineering 5: metal-enhanced fluorescence and plasmon emission "Anal. Biochem. **337**, 171 – 194 (2005).
20. I. Gryczynski, J. Malicka, Z. Gryczynski, J. R. Lakowicz, "Surface plasmon-coupled emission with gold films," J. Phys. Chem. B **108**, 12568 – 12574 (2004).
21. H. Du, R. A. Fuh, J. Li, A. Corkan, J. S. Lindsey, "PhotochemCAD: A computer-aided design and research tool in photochemistry " Photochem. Photobiol. **68**, 141-142 (1998).
22. G. Jones, W. R. Jackson, C.-Y. Choi, W. R. Bergmark, "Solvent effects on emission yield and lifetime for coumarine laser-dyes – requirement for a rotatory decay mechanism," J. Phys. Chem. **89**, 294 – 300 (1985).
23. E.Verhagen, A. L. Tchebotareva, A. Polman, "Erbium luminescence imaging of infrared surface plasmon polaritons," Appl. Phys. Lett. **88**, 121121 (2006).
24. E. Verhagen, L. Kuipers, A. Polman, "Enhanced nonlinear optical effects with a tapered plasmonic waveguide," Nano Lett. **7**, 334 – 337 (2007).
25. F. Lopez-Tejeira, Sergio G. Rodrigo, L. Martin-Moreno, F. J. Garcia-Vidal, E. Devaux, T. W. Ebbesen, J. R. Krenn, P. Radko, S. I. Bozhevolnyi, M. U. Gonzalez, J. C. Weeber, A. Dereux "Efficient unidirectional nanoslit couplers for surface plasmons," Nat. Phys. **3**, 324 – 328 (2007).